\documentclass[pra,twocolumn,superscriptaddress]{revtex4}
\usepackage{blindtext}
\usepackage{amsfonts}
\usepackage{amsmath,amssymb}
\usepackage{graphicx}
\usepackage{subfigure}
\usepackage{float}
\usepackage{bm}
\usepackage{epstopdf}
\usepackage{url}
\usepackage{hyperref}
\usepackage{xcolor}
\hypersetup{
     colorlinks = true,
     linkcolor = blue,
     anchorcolor = blue,
     citecolor = blue,
     filecolor = blue,
     urlcolor = blue
}

\begin{document}

\title{Critical exponents and fluctuations at BEC in a 2D harmonically trapped ideal gas}
\author{M.I. Morales-Amador, V. Romero-Roch\'{\i}n  and R. Paredes} 
\affiliation{Instituto de F\'{\i}sica, Universidad
Nacional Aut\'onoma de M\'exico, Apartado Postal 20-364, M\'exico D.F. 01000, Mexico.}  
\email{rosario@fisica.unam.mx}
\begin{abstract}
The critical properties displayed by an ideal 2D Bose gas trapped in a harmonic potential are determined and characterized in an exact numerical fashion. Beyond thermodynamics, addressed in terms of the global pressure and volume which are the appropriate variables of a fluid confined in a non-uniform harmonic potential, the density-density correlation function is also calculated and the corresponding correlation length is found. Evaluation of all these quantities as Bose-Einstein condensation (BEC) is approached manifest its critical continuous phase transition character. The divergence of the correlation length as the critical temperature is reached, unveils the expected spatial scale invariance proper of a critical transition. The logarithmic singularities of this transition are traced back to the non-analytic behavior of the thermodynamic variables at vanishing chemical potential, which is the onset of BEC. The critical exponents associated with the ideal BEC transition in the 2D inhomogeneous fluid reveals its own universality class.
\end{abstract}

\maketitle

\section{Introduction}

 Finite temperature phase transitions have remained until the present as one of the most intriguing and challenging phenomena explored across diverse fields in physics, as they exhibit the essence of collective behavior \cite{Landau,Stanley}. Mostly, second order or critical phase transitions, due to their large density fluctuations near the critical points, allow for understanding them in terms of universality classes whose theoretical description can be cast in terms of classical fields, with the renormalization group being the main tool \cite{Wilson,Ma,Amit}. Within these, and of relevance to the present study, the origin of critical phase transitions in superfluid \cite{Fisher,Pethick,Campostrini,Burovski,Lipa}, superconductivity \cite{Abrikosov,Fetter,Annett} and Bose-Einstein condensation \cite{Landau, Reif, Huang, Pathria} resides not only in the collective but also in the statistical quantum nature of the atomic constituents of the system. While interatomic interactions are fundamental for most of the peculiarities of the diverse universality classes, the critical properties of ideal Bose-Einstein condensation emerge from purely intrinsic quantum statistical exchange effects. 

In accord with the prescription of the Mermin-Wagner theorem \cite{Mermin-Wagner,Hohenberg} that denies the existence of long range order in homogeneous systems with dimension equal or lower than two, ideal Bose-Einstein condensation (BEC) is forbidden in homogeneous two dimensional space, being this fact a direct consequence of the divergence of the Bose function $g_{1/2}(\mu/k_BT)$ as the chemical potential approaches zero $\mu \to 0^-$, see below.  It should be mentioned that it has been argued that such a restriction requires of the thermodynamic limit and that it would not hold for a finite number of particles \cite{Cho}. However, the occurrence of 2D BEC in harmonically trapped ideal \cite{Bagnato,Mullin,Romero-Rochin,Brange} and interacting atomic gases \cite{Kruger,Rajagopal,Tiesinga,Dalibard} is not outlawed. Here is suited to mention that interacting Bose gases in 2D undergo a BKT phase transition \cite{Hadzibabic,Fletcher,Sunami,Singh}. As a signature of this continuous transition one can mention the measurement of the correlation length of a 3D gas in $^{87}$Rb atoms confined in a 3D harmonic trap by Donner et al. \cite{Donner}, where they found the critical exponent $\nu \approx 0.67 \pm 0.13$ in very good agreement with the 3D XY-model \cite{Campostrini,Burovski}, the universality class of superfluid $^4$He \cite{Lipa}. Hung et al. \cite{Hung} have also reported on the measurement of density-density correlation function and the static structure factor in a 2D gas of  $^{133}$Cs atoms. Hence, in view of the experimental facilities of manipulating, not just interactions, but also external confinements settings, as well as dimensional degrees of freedom \cite{Cornell, Ketterle, Dalfovo, Jin, Gerbier}, a thorough analysis of the BEC phenomenon remains as a current topic of interest. In this article we focus our attention in the occurrence of the condensation transition in an ideal gas of Bose atoms trapped by a harmonic potential in two dimensions, with special emphasis on the critical characteristics of the thermodynamic properties and of the density fluctuations. 

The main thermodynamic properties are here obtained using the traditional Grand Canonical ensemble. The proper identification of intensive and extensive thermodynamic mechanical variables pertaining to a harmonic trap, and that may be called global pressure ${\cal P}$ and global volume ${\cal V}$ \cite{Romero-Rochin,Romero-Rochin3,Sandoval-Figueroa}, allows us to deal with the appropriate equation of state ${\cal P} = {\cal P}(N/{\cal V},T)$. In the same way, we can thus determine the usual fluid susceptibilities, namely, heat capacities $C_{\cal V}$ and $C_{\cal P}$ at constant volume and pressure, respectively, and the isothermal compressibility $\kappa_T$. After the usual determination of the Bose-Einstein condensation in the system, we then study in detail the critical behavior of the previous quantities in the vicinity of the BEC transition. In particular, we obtain the corresponding critical exponents of those quantities, being $\alpha_c = -1$, $\gamma_c = 0$ (purely logarithmic) and $\delta_c = 1$.  For sure, we can obtain these results since the full quantum statistical problem can be solved analytically, with direct numerical evaluation of some quantities, but without appealing to scaling or renormalization group like arguments. As we shall see, besides the expected algebraic dependences in terms of those exponents, there appear logarithmic corrections \cite{Wegner,Adler,Kenna}, not arising from finite size effects. The precise values of those exponents, concomitant to the accompanying logarithmic corrections, indicate that ideal BEC in a 2D harmonic trap has its own critical universality class, different of both ideal 3D BEC in uniform and in harmonic traps. The final and crucial signature that allows us to identify the critical transition is the behavior of the density-density correlation function in the vicinity of the temperature at which condensation happens, in the thermodynamic limit. Larger and larger density fluctuations are hence observed in the inhomogeneous environment created by the harmonic trap, as the thermodynamic limit nears. An important finding of our study is the determination of the correlation length $\xi$ which, indeed, diverges as Bose-Einstein condensation is established. We also find the corresponding critical exponent $\nu_c = 1/2$ of the correlation length, with logarithmic corrections as well.

The article is organized as follows. In section II we study the thermodynamics of the ideal Bose gas confined in a 2D isotropic harmonic trap, revising the natural emergence of the global pressure and volume and, then, the BEC phenomenon is determined with attention to the critical properties arising from the equation of state. In section III we analyze the mentioned susceptibilities and their critical properties as well. Section IV is devoted to the density-density correlation function which allows for observing density critical fluctuations and, as mentioned, we find the corresponding correlation length. We conclude with some final remarks.

\section{Equation of state of an ideal Bose gas in a harmonic trap in 2D}
\label{Model}

The Grand Potential for a gas of bosons of mass $m$ and spin $s = 0$ confined in a 2D isotropic harmonic trap of frequency $\omega$, for a given temperature $T$ and chemical potential $\mu$, is given by \cite{Landau, Reif, Huang, Pathria},
\begin{equation}
\Omega(\mu, T, \omega)=k_BT \sum_{\bf m} \ln \left( 1 - e^{-\beta \hbar \omega (m_x+m_y) +\alpha} \right)
\label{GP}
\end{equation}
where $\beta=1/k_B T$, $\alpha=\mu/k_B T$ and ${\bf m} = (m_x,m_y)$ are the 2D harmonic oscillator quantum numbers, with
\begin{equation}
\sum_{\bf m} = \sum_{m_x = 0}^\infty \sum_{m_y = 0}^\infty \>.
\end{equation}
As discussed in \cite{Romero-Rochin3,Sandoval-Figueroa}, and further addressed below, the thermodynamic limit for this type of confinement is $N \to \infty$ and $\omega \to 0$, keeping $N\omega^2 =$ constant;  this allows us to introduce a continuous density of states and transform the sums into integrals over the single particle energy,
\begin{equation}
\sum_{\bf m} \to  \int_0^\infty \frac{\epsilon}{(\hbar \omega)^2} d\epsilon \>.\label{DOS}
\end{equation}
 Then, the Gran Potential adopts the form,
\begin{equation}
\Omega(\mu, T, \omega)= -k_B T \left ( \frac{k_B T}{\hbar \omega} \right)^2 g_3(\alpha) \>,\label{Ome0}
\end{equation}
where we have introduced the Bose function
\begin{equation}
g_z(\alpha) =\frac{1}{\Gamma(n)} \int_0^\infty dx  \frac{x^{z-1}}{e^{x-\alpha}-1} \>,
\end{equation}
with $\alpha = \mu/k_BT$. 

Since the grand potential $\Omega$ must be an extensive thermodynamic quantity \cite{Landau}, from (\ref{Ome0}) one observes that the only thermodynamic quantity in its right-hand side that can have such a property is $1/\omega^2$. This indicates that one can define a generalized thermodynamic variable, called ``global volume'' ${\cal V} = 1/\omega^2$, as it plays the same role of the usual volume $V$ of a confining vessel of rigid walls. Indeed, the smaller $\omega$ is, the larger the physical volume occupied by the gas, for a given number of particles $N$ and temperature $T$. As a matter of fact, such a property has already been used in the thermodynamic limit above as $N \to \infty$, ${\cal V}  \to \infty$, with $N/{\cal V} =$ constant, for obtaining (\ref{Ome0}). This identification allows us to write $d\Omega = - SdT - N d\mu - {\cal P}d{\cal V}$, thus introducing a ``global pressure'' ${\cal P}$, such that $\Omega = - {\cal PV}$. One can notice that although ${\cal V}$ and ${\cal P}$ do not have the standard units of volume and pressure, the product $\Omega= -{\cal P} {\cal V}$ does have the correct units of energy. 
The proper identification of these variables, as the correct thermodynamic mechanical variables for harmonic traps, has thoroughly been discussed in Refs. \cite{Romero-Rochin, Romero-Rochin3, Sandoval-Figueroa} and it has already been used in the analysis of real experiments in ultracold gases, see e.g.  \cite{Vander-banda,Vander-banda-2}. 

With the previous identifications one can find the number of particles $N$, the energy $E$, the entropy $S$ and the global pressure ${\cal P}$,
\begin{eqnarray}
N&=& \left(\frac{k_B T}{\hbar \omega} \right)^2 g_2(\alpha) \>, \label{N} \\
E& =& 2k_B T \left(\frac{k_B T}{\hbar \omega} \right)^2 g_3(\alpha) \>,  \label{E} \\
S& =& k_B \left(\frac{k_B T}{\hbar \omega} \right)^2 \left[3 g_3(\alpha)- \alpha g_2(\alpha) \right] \>, \label{S} 
\end{eqnarray}
and

\begin{equation}
{\cal P} = k_B T \left( \frac{k_B T}{ \hbar} \right)^2 g_3(\alpha) \>.\label{P}
\end{equation}

While it is very well known that a gas confined in a 2D box of rigid walls does not present the phenomenon of BEC, the present gas does so. The occurrence of the phenomenon can be traced back to the single-particle density of states, scaling as $\epsilon^{0} = 1$ in the 2D box and here, as shown in (\ref{DOS}), as $\epsilon^1$. BEC can thus be found to occur by looking at the expression for the number of particles $N$ in (\ref{N}). In the 2D box case $N \sim g_{1/2}(\mu/k_BT)$ a quantity that diverges as $\mu \to 0$. However, for the harmonic confinement, as one can easily verify, say, for fixed global density $N/{\cal V} = N\omega^2$, there exists a finite critical temperature $T_c$ at which the chemical potential vanishes, $\mu = 0$, with the corresponding Bose function remaining finite, $g_2(0) = \zeta(2)$, the zeta function of 2. Explicitly, this condition is, see (\ref{N}),
\begin{equation}
\frac{N}{\cal V} = \left(\frac{k_B T}{\hbar} \right)^2 \zeta(2) \>,\label{Tcrit}
\end{equation}
which yields a critical temperature if the density $N/{\cal V}$ is fixed, or conversely, a critical density if the temperature is the given quantity. As in the textbook descriptions of BEC, in the thermodynamic limit here we find that above condensation, $T > T_c$ or alternatively $\mu < 0$, the condensate fraction $N_0/N$ is zero, with $N_0$ the number of particles in the one-particle ground state $\epsilon = 0$. For $T < T_c$, the chemical potential remains zero, $\mu = 0$, and $N_0$ is then macroscopic. Using (\ref{N}) and  (\ref{Tcrit}), one can summarize BEC as,
\begin{equation}
\frac{N_0}{N} = \left\{
\begin{array}{ccc}
0 & {\rm if} & T \ge T_c \\
1 - \left(\frac{T}{T_c}\right)^2 & {\rm if} & T \le T_c
\end{array} \right. \>. \label{conden-frac}
\end{equation}
One should keep in mind that the critical temperature $T_c$ depends on the current global density of the gas $N/{\cal V}$, as given by (\ref{Tcrit}).

The above thermodynamic expressions, as arising from the Grand Canonical Ensemble, are expressed in terms of the independent variables $({\cal V},T,\mu)$. However, typically, the equation of state is expressed in terms of the canonical variables $(N,{\cal V},T)$, namely, ${\cal P}= {\cal P}(N/{\cal V}, T )$. The change of variables $\mu$ by $N/{\cal V}$ can be numerically performed by inverting equation (\ref{N}). However, it is of interest to present hybrid expressions for the pressure and the energy in the following form, by combining (\ref{N}), (\ref{E}) and (\ref{P}), 
\begin{equation}
{\cal P} = \frac{N k_B T}{{\cal V}} \frac{g_3(\alpha)}{g_2(\alpha)},
\label{P2}
\end{equation}
and
\begin{equation}
E=2 N k_B T \frac{g_3(\alpha)}{g_2(\alpha)}.
\end{equation}
Since the classical limit is obtained \cite{Landau} when $\mu \to -\infty$ and all Bose functions are $g_z(\alpha) \approx e^{\alpha}$ in such a limit, one obtains ${\cal PV} \approx Nk_BT$, the equation of state of an ideal classical gas, and $E \approx 2 Nk_BT$, the restoration of the classical equipartition theorem.

\begin{figure}[h]
\begin{center}
\includegraphics[width=8.5cm, height=5.7cm]{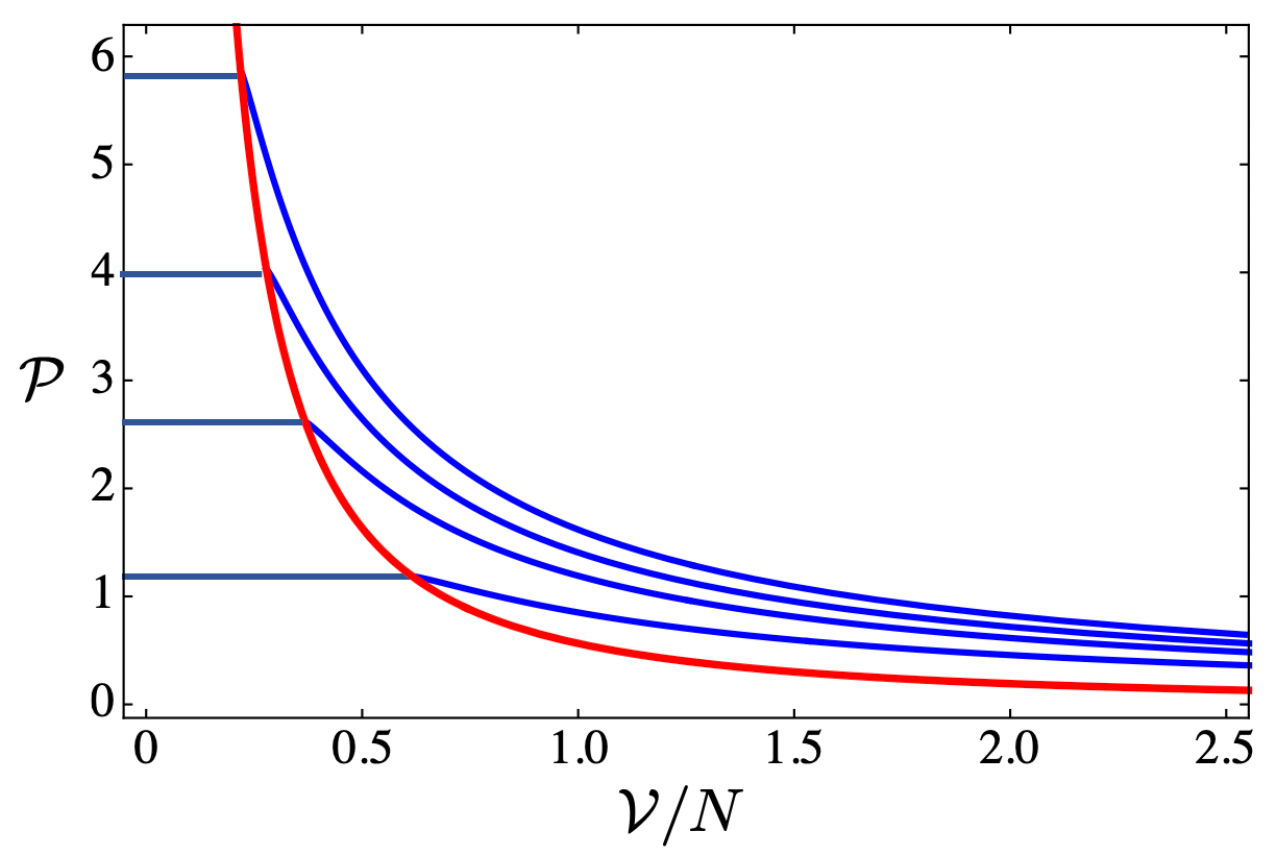}
\end{center}
\caption{(Color on line) Isotherms of the equation of state ${\cal P}(N/{\cal V}, T)$. Below the transition line the harmonic pressure remains constant.}
\label{Fig1}
\end{figure}

\begin{figure}[h]
\begin{center}
\includegraphics[width=8.5cm, height=5.7cm]{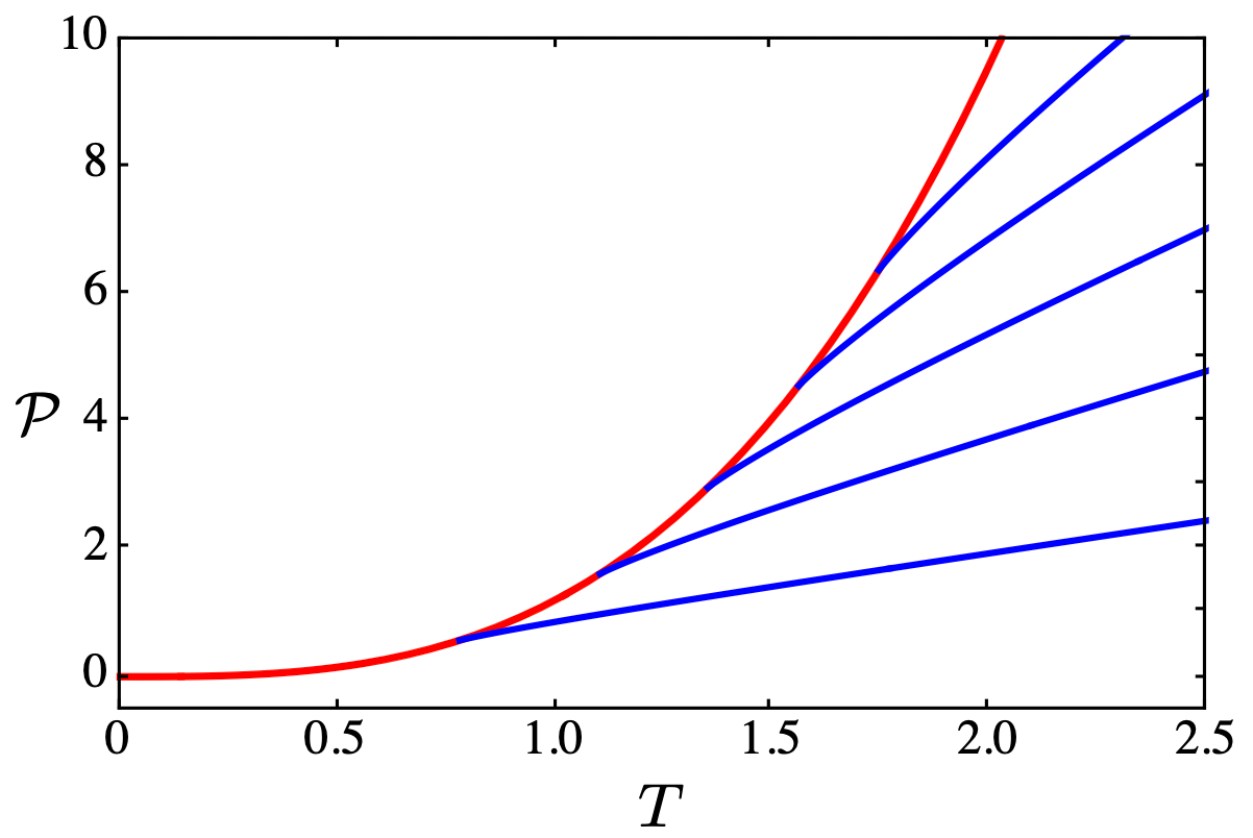}
\end{center}
\caption{Isochores of the equation of state ${\cal P}(N/{\cal V}, T)$. All the isochores merge at the transition line which is indicated with continuous red line.}
\label{Fig2}
\end{figure}

In general, the equation of state ${\cal P}= {\cal P}(N/{\cal V}, T )$ shows the BEC phenomenon similarly to the textbook case of a gas in a 3D box \cite{Pathria} and the gas confined in a 3D harmonic trap \cite{Romero-Rochin}. In figure \ref{Fig1} we plot isotherms in a ${\cal P}-N/{\cal V}$ diagram, with the critical curve scaling as ${\cal P} \sim (N/{\cal V})^{3/2}$. Similarly, in figure \ref{Fig2} we show isochores, namely, curves of constant global density $N/{\cal V}$, in a ${\cal P}-T$ diagram. The corresponding critical curve scales as ${\cal P} \sim T^3$. 

As far as the elucidation of the thermodynamic properties near the critical curve $\mu = 0$, it is useful to recall the asymptotic series of the integer $n \ge 1$ Bose functions $g_n(\alpha)$ for $|\alpha |< 1$, \cite{Wiki}
\begin{equation}
g_n(\alpha) = \frac{\alpha^{n-1}}{(n-1)!}\left(H_{n-1}-\ln(-\alpha)\right) + \sum_{k=0,k\ne n-1}^\infty \frac{\zeta(n-k)}{k!} \alpha^k \>, \label{asint}
\end{equation}
where $H_m = \sum_{k=1}^m(1/k)$ and $H_0 = 0$. We see right away that the first term is non-analytic at $\alpha = 0$, indicating the critical nature of the BEC phase transition. 

With expression (\ref{N}) for the number of particles $N$ we can enquire about a usual critical property, which is the relationship between the pressure and the volume as the critical line is approached, along an isotherm of constant temperature $T$. In this way, for $|\mu/k_BT| \ll 1$, using (\ref{N}), (\ref{P}) and (\ref{asint}), the global density $\rho^G \equiv N/{\cal V}$ and the pressure behave, for sufficiently small $|\alpha|$, as 
\begin{eqnarray}
\frac{\rho^G - \rho^G_c}{\rho^G_c}&\approx& -\frac{\alpha}{\zeta(2)} \ln(-\alpha) \\
\frac{{\cal P} - {\cal P}_c}{{\cal P}_c}&\approx& \frac{\zeta(2)}{\zeta(3)} \alpha \>,
\end{eqnarray}
where $\rho^G_c$ and ${\cal P}_c$ are the critical global density and pressure at $\mu = 0$. Combination of  these expressions yields the sought for result,
\begin{equation}
\frac{{\cal P}_c - {\cal P}}{{\cal P}_c} \ln\left(\frac{\zeta(3)}{\zeta(2)}\frac{{\cal P}_c - {\cal P}}{{\cal P}_c}\right) \approx \frac{\zeta^2(2)}{\zeta(3)} \frac{\rho^G - \rho^G_c}{\rho^G_c} \>.
\end{equation}
Appealing to the general theory of critical phenomena \cite{Stanley,Ma,Amit}, one expects a scaling relationship of the form ${\cal P}-{\cal P}_c \sim (\rho^G - \rho^G_c)^{\delta_c}$ near a critical point, thus defining the critical exponent $\delta_c$. From the above expression, we find that we can assign the exponent $\delta_c = 1$ but, very importantly, with trascendental logarithmic corrections. In the final section we will return to the validity and issues of the present identification of the critical exponents and to the discussion of the logarithmic corrections.

\section{Susceptibilities}

As BEC is a critical phase transition, the disclosure of the characteristic and unique properties of the critical phenomenon at hand can be best analyzed by studying the behavior of both the susceptibilities and the particle density critical fluctuations in the vicinity of the transition. We leave the study of the latter for the next section and here our attention is devoted to the pure thermodynamic behavior of the heat capacities at constant global volume $C_{\cal V}$ and at constant global pressure $C_{\cal P}$ and the isothermal compressibility $\kappa_T$  
given by,
\begin{eqnarray}
C_{\cal V} &=& T \left(\frac{\partial S}{\partial T}\right)_{{\cal V},N}  \label{CV}\\
C_{\cal P} &=& T \left(\frac{\partial S}{\partial T}\right)_{{\cal P},N} \label{CP} \\
\kappa_T & = & -\frac{1}{\cal V} \left(\frac{\partial {\cal V}}{\partial {\cal P}}\right)_{T,N} = \frac{\cal V}{N^2} \left(\frac{\partial N}{\partial \mu}\right)_{T,{\cal V}} \label{KT}
\end{eqnarray}
Special attention must be paid to determine these quantities since they are all considered at constant number of particles $N$ and this is meaningful above condensation only since, in such a case, $N$ is an independent thermodynamic variable that determines the state of equilibrium of the gas. Hence, for $T > T_c$, one can calculate the above susceptibilities,  showing their explicit expressions below in terms of $\mu$ and $T$. However, below BEC or $T \le T_c$, the chemical potential reaches its maximum value $\mu=0$ and remains so, thus obliterating $N$ as an independent thermodynamic variable that determines the state. Indeed, below condensation, the fraction of particles either or not in the condensate are functions of the temperature $T$, as indicated by (\ref{conden-frac}). In other words, the Grand Potential is a function of two variables only, $T$ and $\omega$,
\begin{equation}
\Omega(\mu, T, \omega)= -k_B T \left ( \frac{k_B T}{\hbar \omega} \right)^2 \zeta(3) \>\>\>\>\> T \le T_c .\label{Ome1}
\end{equation}
Accordingly, one can find the entropy and the pressure only, the number of particles and the chemical potential being no longer independent thermodynamic variables that specify the state,
\begin{equation}
S({\cal V},T)= 3 k_B  \left( \frac{ k_B T}{\hbar \omega} \right)^2 \zeta(3) \>\>\>\>\> T \le T_c,\label{SBEC}
\end{equation}
and
\begin{equation}
{\cal P}=  k_B T \left( \frac{k_B T }{\hbar } \right)^2 \zeta(3) \>\>\>\>\> T \le T_c\>.\label{PBEC}
\end{equation}
The last equation indicates that the pressure ${\cal P}$ depends only on $T$, no longer on $N/{\cal V}$. This has the interesting result that $C_{\cal P}$, $\kappa_T$ and $\alpha_T$ are no longer meaningful susceptibilities, not only from a mathematical point of view, but from the physical one as well. That is, if ${\cal P}$ is constant, so is $T$ and viceversa, and therefore we cannot physically enquire as to the amount of heat needed to increase the temperature if the pressure ${\cal P}$ remains constant. Analogous comments apply to the meaning of $\kappa_T$.
This should not be surprising, as the same comment applies to black-body radiation in a cavity of volume $V$, where the free energy depends on $T$ and the volume $V$ of the cavity only, namely, $F = F(V,T)$ \cite{Landau,Huang,Pathria}; in such a physical problem, one never worries about its isothermal compressibility, for example, as it is not a physically meaningful quantity. We readdress this issue in the following section.

\begin{figure}[h]
\begin{center}
\includegraphics[width=8.5cm, height=5.7cm]{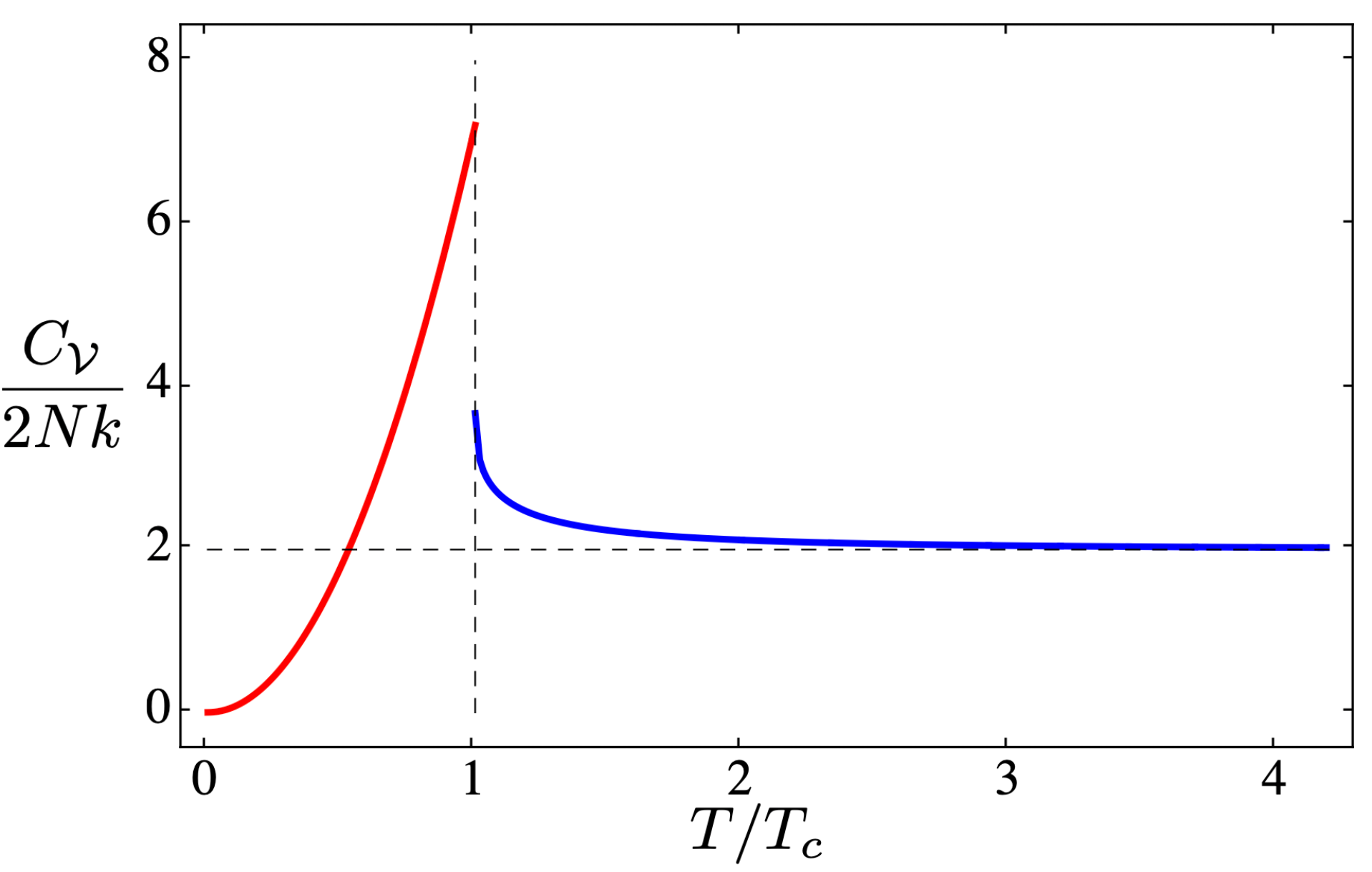}
\end{center}
\caption{Heat capacity at constant harmonic volume, $C_{{\cal V}}$, as a function of temperature $T/T_c$. Note the discontinuity at $T_c$ and the classical equipartition value $C_{{\cal V}} \to 2N k_B$  at high temperature.}
\label{Fig3}
\end{figure}

With the above remarks we can now present the behavior of the susceptibilities. We start by considering $C_{\cal V}$,  finding,
\begin{equation}
C_{\cal V}
= \left\{
\begin{array}{ccc}
6 k_B  \left( \frac{k_B T }{\hbar \omega} \right)^2 \zeta(3) & {\rm for} & T < T_c\\
& & \\
2 N k_B  \left[ 3 \frac{g_3(\alpha)}{g_2(\alpha)} - 2\frac{g_2(\alpha)}{g_1(\alpha)} \right]  & {\rm for} & T > T_c
\end{array}\right. \label{CV1}
\end{equation}
In figure (\ref{Fig3}) it is shown the behavior of $C_{{\cal V}}$ as a function of temperature $T$, for a given value of the global density $N/{\cal V}$. For $T > T_c$ one must combine expressions (\ref{N}) and (\ref{CV1}) to obtain the curve shown in the figure. For $T < T_c$ and ${\cal V}$ fixed, one can indeed enquire about the amount of heat needed to change the temperature in one unit, namely, of the value of $C_{\cal V}$ and, then, one can use the entropy expression (\ref{SBEC}) to calculate it. Two noticeable features deserve attention. On the one side the discontinuity  at $T=T_c$.  As in all systems that show ideal BEC the behavior\cite{Landau,Huang,Pathria,Romero-Rochin} of $C_{\cal V}$ is reminiscent of the lambda transition in superfluid $^{4}$He \cite{Lipa}, where the heat capacity neither diverges but becomes sharply peaked only.
On the other side, one can also realize how the classical behavior for high temperatures $T \gg T_c$, $C_{{\cal V}} \to 2N k_B$ is reestablished, obtaining the Dulong-Petit law. 

Regarding the critical behavior of $ C_{{\cal V}}$ as BEC is approached, we can use first the asymptotic series (\ref{asint}) of the Bose functions in (\ref{CV1}) for $T > T_c$, finding
\begin{equation}
C_{\cal V} \approx C_{\cal V}^c \left(1 + \frac{\zeta^2(2)}{3}\frac{\mu}{k_BT_c}\right)\>,\label{CVcrit}
\end{equation}
where $C_{\cal V}^c$ is the critical value of the heat capacity as $T \to T^+$. We would like now to express the critical behavior not in terms of $\mu$, but in terms of the approach 
of the temperature to its critical value, that is, in terms of $T-T_c$. For this, we can use expressions (\ref{N}) and (\ref{Tcrit}) for $N$ and $T_c$ and the asymptotic expression for $g_2(\alpha)$, see (\ref{asint}); one finds, near BEC
\begin{equation}
\frac{T-T_c}{T_c} \approx - \frac{1}{\zeta(2)} \left| \frac{\mu}{kT_c}\right|  \ln \left| \frac{\mu}{kT_c}\right| \>.\label{muTc}
\end{equation}
This is also a trascendental relationship, certainly showing that as $\mu$ vanishes so does $T-T_c$, but with logarithm corrections. Critical phenomena theory \cite{Stanley,Ma,Amit} indicates that the behavior of the heat capacity at constant volume should scale as $C_{\cal V} \sim |T-T_c|^{-\alpha_c}$ near a critical point. Hence, from the above expressions (\ref{CVcrit}) and (\ref{muTc}), we can assign the critical exponent $\alpha_c = -1$, with logarithmic corrections. For $T \to T_c^{-}$, that is approaching BEC from the condensate region, using (\ref{CV1}) for $T < T_c$, one also finds the exponent $\alpha_c = -1$ but with no logarithmic corrections. It is interesting to recall that the critical exponent $\alpha_c$ at the superfluid $^4$He transition \cite{Lipa} is also negative, as well as in BEC in a 3D box \cite{Reyes-Ayala}.

\begin{figure}[h]
\begin{center}
\includegraphics[width=8.5cm, height=5.7cm]{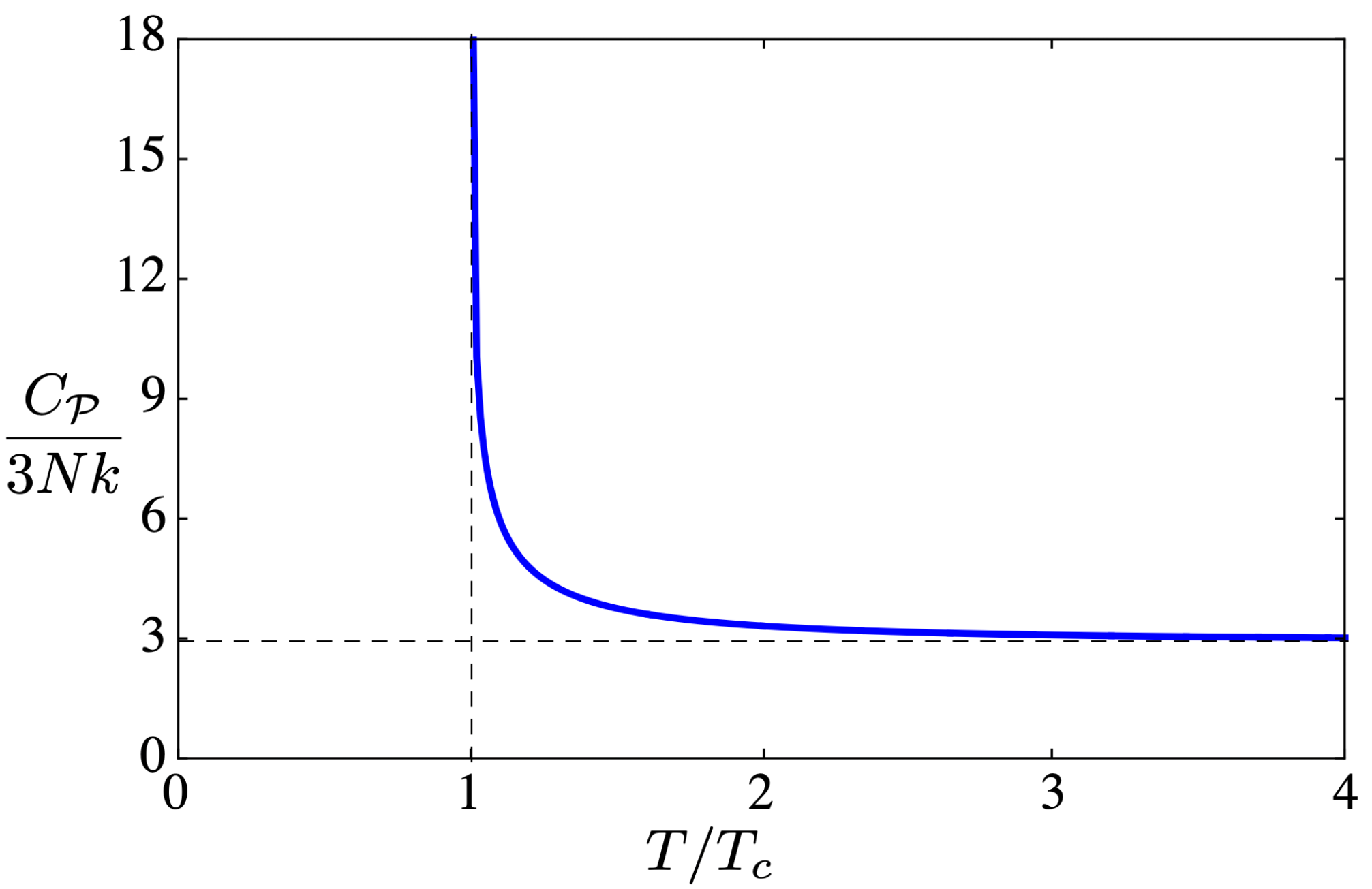}
\end{center}
\caption{Heat capacity at constant harmonic pressure, $C_{{\cal P}}$ , as a function of temperature $T/T_c$. Note the discontinuity at $T_c $ and the classical equipartition value $C_{{\cal P}} \to 3N k_B$  at high temperature.}
\label{Fig4}
\end{figure}

 In the case of the heat capacity at constant global pressure one finds the following expression for $T > T_c$,
\begin{equation}
C_{{\cal P}} = 3 N k_B \frac{g_1(\alpha) g_3(\alpha))}{g_2^2 (\alpha)} \left[ 3 \frac{g_3(\alpha)}{g_2(\alpha)} - 2\frac{g_2(\alpha)}{g_1(\alpha)} \right] \>,
\label{CP1}
\end{equation}
and in figure (\ref{Fig4}) we show it as a function of temperature $T$, for a fixed value of the global density $N/{\cal V}$. Again, one can use (\ref{N}) to calculate the curve shown in the figure. We first observe that in the classical limit $\mu \to -\infty$, $C_{{\cal P}} \approx 3 N k_B$, yielding the ratio $C_{{\cal P}}/C_{{\cal V}} \to 3/2$, as expected for 2D harmonic systems. However,  we highlight the very interesting behavior, also shared by $\kappa_T$ as discussed below, that $C_{{\cal P}}$ diverges as $T \to T_c^{+}$. The origin of this divergence is due to the presence of the Bose function $g_1(\alpha)$  in the right-hand side of (\ref{CP1}), which shows a logarithmic divergence as $\alpha \to 0^{-}$, as it can be seen from (\ref{asint}). Using such an expansion, one finds the critical behavior of $C_{{\cal P}}$ at constant global density,
\begin{equation}
C_{{\cal P}} \approx - 9 N k_B \frac{\zeta^2(3)}{\zeta^3(2)} \ln \left| \frac{\mu}{kT_c}\right|
\end{equation}
showing a logarithm divergence in terms of the chemical potential. By means of (\ref{muTc}) we also can also conclude that  the specific heat at constant global pressure diverges logarithmically with $(T-T_c)/T_c$, with further logarithmic corrections. The corresponding critical exponent will be seen below to be zero.

\begin{figure}[h]
\begin{center}
\includegraphics[width=8.5cm, height=5.7cm]{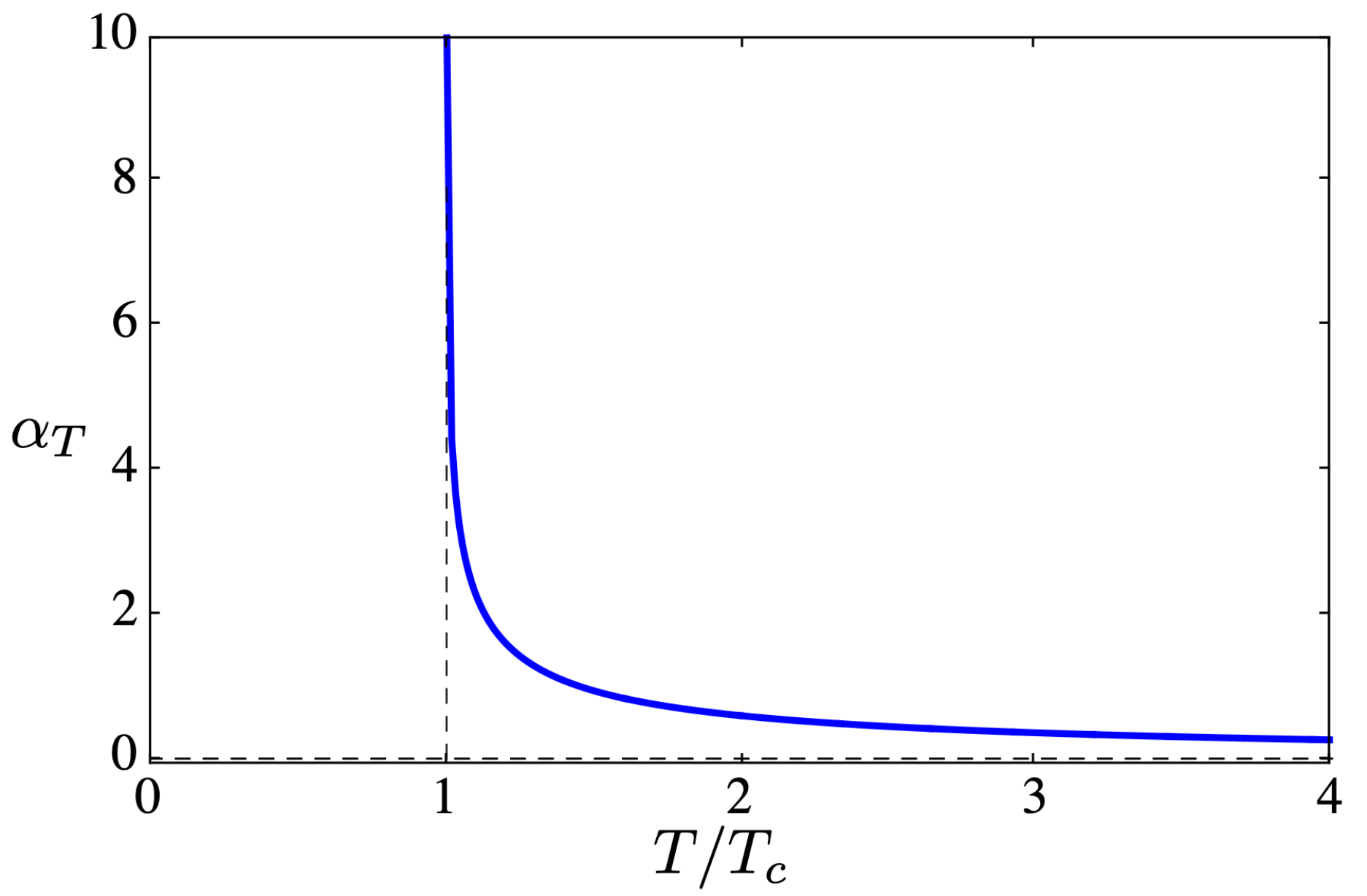}
\end{center}
\caption{Thermal expansion coefficient $\alpha_T$, as a function of temperature $T/T_c$. Note the divergence at $T_c$.}
\label{Fig5}
\end{figure}

We now turn our attention to the  isothermal compressibility $\kappa_T$, defined in (\ref{KT}) and explicitly given by,

\begin{equation}
\kappa_T= \frac{{\cal V}}{N k_B T} \frac{g_1(\alpha)}{g_2(\alpha)} \>,\label{KT1}
\end{equation}
for $T \ge T_c$. Figure (\ref{Fig5}) shows $\kappa_T$ as a function of $T$ for constant global density $N/{\cal V}$. In the classical limit, one reaches the ideal gas Curie-like law $\kappa_T \sim 1/T$. And, as expected from general thermodynamic considerations near critical points \cite{Landau,Huang,Pathria}, and corroborated by the presence of the Bose function $g_1(\alpha)$, the isothermal compressibility shares the same critical behavior as the specific heat at constant pressure, that is, it diverges logarithmically, with further logarithm corrections, as $T \to T_c$,
\begin{eqnarray}
\kappa_T &\approx & - \frac{{\cal V}}{\zeta(2) N k_B T_c} \ln \left| \frac{\mu}{kT_c}\right| \nonumber \\
&\approx & - \frac{{\cal V}}{\zeta(2) N k_B T_c} \ln \left| \frac{T-T_c}{T_c}\right| \>,
\end{eqnarray}
where the second line follows from (\ref{muTc}) at lowest order. This indicates that the critical exponent $\gamma_c$, in the usual expression near the critical point \cite{Stanley,Ma,Amit}, $\kappa_T \sim |T-T_c|^{-\gamma_c}$, must be taken as $\gamma_c = 0$. It is interesting to contrast this critical behavior of the isothermal compressibility with other systems that show ideal BE condensation. For instance, for the gas in a 3D box \cite{Reyes-Ayala,Gunton}, the isothermal compressibility diverges with exponent $\gamma = 1$, while for the gas in a 3D harmonic potential \cite{Romero-Rochin} the isothermal compressibility does not diverge at all, reaching a finite value at condensation.
As one should expect from general considerations of critical phenomena, a divergent compressibility should be associated with critical density fluctuations in physical real space. This is the subject of the following section, where we will show that the density-density correlation length indeed diverges as BEC is approached.

\section{Density-Density correlations and critical fluctuations}
\label{Correlations}

The density-density correlation function, a many-body quantity that goes beyond the realm of pure thermodynamics and based on the underlying atomic nature of a fluid, yields precious information about particle density fluctuations \cite{Bors,Hohenberg}. In this regard, one of the cornerstones of our understanding of critical phenomena is the fact that, as a critical point is approached, typically the density fluctuations grow with no bound. This can be directly observed by monitoring the growth of the correlation length $\xi$ of such fluctuations as the system nears the critical point. In the BEC phenomenon, similarly to the superfluid and superconducting transitions, there is a curve of critical points given by $T_c$, as in (\ref{Tcrit}), with one critical temperature for a given thermodynamic global density $N\omega^2$, at zero chemical potential. Thus, in order to study the density fluctuations and find the corresponding correlation length, we start with the most general expression for the density correlation function for ideal bosons placed at two spatial points in the fluid, one at ${\bf r}$ and the other lying at ${\bf r'}$ \cite{Fetter}: 
\begin{equation}
C({\bf r}, {\bf r'})= \langle {\hat \rho}({\bf r}) {\hat \rho}({\bf r'})\rangle - \langle{\hat \rho}({\bf r}) \rangle \langle{\hat \rho}({\bf r'})\rangle,
\label{correl}
\end{equation}
where ${\hat \rho}({\bf r})= {\hat \Psi}^\dagger({\bf r}) {\hat \Psi}({\bf r})$ is the particle density operator, given in terms of the field operators ${\hat \Psi}^\dagger({\bf r})$ and ${\hat \Psi}({\bf r})$ that create and annihilate particles at position ${\bf r}$. These operators can be written in terms of the one-particle wavefunctions $\phi_{{\bf m}}({\bf r})$ of a 2D harmonic potential, 
\begin{equation}
{\hat \Psi}({\bf r}) = \sum_{\bf m} \phi_{{\bf m}}({\bf r}) \hat a_{\bf m} \>,
\end{equation}
with $\hat a_{\bf m}$ the annihilation operator of a particle in the state $\phi_{{\bf m}}({\bf r})$, with ${\bf m}=(m_x,m_y)$  the label to identify the states in cartesian coordinates, as stated in (\ref{GP}) above.

By taking the average of the operators in (\ref{correl}), using the grand canonical equilibrium density matrix, one finds a closed expression for the correlation function, see Appendix for details,  
\begin{equation}
 C( {\bf r} ,{\bf r'})= \rho({\bf r})\delta^2({\bf r}-{\bf r}')+ \left| \sum_{{\bf m}} \phi_{{\bf m}}^*({\bf r})\phi_{\bf m}({\bf r}')\bar{n}_{{\bf m}} \right | ^2
 \label{eqncorrel}
 \end{equation}
\noindent
where $\bar{n}_{\bf m}$ is the average occupation number of bosons in state ${\bf m}$, namely, the Bose-Einstein distribution,
\begin{equation}
\bar{n}_{\bf m} = \frac{1}{e^{\beta(\hbar \omega(m_x+m_y +1)- \mu})-1} \>.
\end{equation}
 
 Given the fact that we are dealing with an isotropic harmonic confinement, polar coordinates are most suitable than cartesian to calculate $C({\bf r},{\bf r'})$ and, while this function cannot be isotropic in $|{\bf r} - {\bf r}^\prime|$, it is particularly simpler when the positions of particles are such that one is located at the center of the harmonic trap and the other in an arbitrary direction placed at a distance $r$ from the origin, thus making $C({\bf r},{\bf r'}) = C(|{\bf r}-{\bf r'}|)$, spherically symmetric. Then, consideration of the wave function $\phi_{\bf m}({\bf r})$ expressed in polar coordinates, see the Appendix for details, leads us to write a concise expression for the density-density correlation of particles in positions $|{\bf r}|=0$ and $|{\bf r'}|=r$. This expression reads, ignoring the delta function in (\ref{eqncorrel}), as
\begin{equation}
C(r)= \frac{4}{\pi^2 \Delta^4}e^{- r^2/\Delta^2} \left | \sum_{n =0} ^{\infty} L_{n}( r^2/\Delta^2) \frac{1}{e^{\beta[ \hbar \omega (2 n +1)-\mu]} -1} \right |^2 \>, \label{cr}
\end{equation}
where we have identified the usual length of a harmonic oscillator, $\Delta^2= \hbar/m\omega$, $L_{n}( x)$ is the Laguerre function and $n$ are natural numbers associated to the 2D oscillator eigenenergies $\epsilon_n=\hbar \omega (2 n +1)$. In this last expression for the energy levels we have already taken into account that the only contribution of the angular momentum quantum numbers to the full sum in (\ref{eqncorrel}) is $M = 0$, see the Appendix.


The correlation function $C(r)$ above is amenable to a numerical evaluation. However, before presenting our conclusions a brief discussion on such a numerical evaluation is in order. First, we recall that the thermodynamic limit in the grand canonical ensemble requires to take the limit $\omega^2 \to 0$. Therefore, since the correlation function $C(r)$ is an intensive quantity, it must converge to a finite function of $r$, for given $T$ and $\mu$, in such a limit and, very importantly, it must be {\it independent} of the frequency $\omega$. This may appear contradictory to expression (\ref{cr}) of $C(r)$ since it appears to explicitly depend on $\omega$.  As explicit numerical calculations as described below corroborate, $C(r)$ is independent of $\omega$ in the thermodynamic limit.
Incidentally, the convergence to its limiting values further validates the identification of ${\cal V} = 1/\omega^2$ as the correct thermodynamic extensive ``volume'' of the gas. Thus, the way we proceed to evaluate $C(r)$ is, first, to keep the global density $N/{\cal V}$ constant allowing for an identification of a precise value of the critical temperature $T_c$, see (\ref{Tcrit}), with which we can define dimensionless quantities $\tilde T = T/T_c$, $\tilde \omega = \hbar\omega/k_BT_c$, $\tilde \mu = \mu/k_BT_c$, $\tilde r = r/(\hbar^2/mk_BT_c)^{1/2}$ and $\tilde C = (\hbar^2/mk_BT_c)^2 C$. We further proceed by then fixing the  chemical potential $\tilde \mu$, which correspondingly yields a temperature $\tilde T$ by means of (\ref{Tcrit}). Then, systematically reducing the value of the 
the dimensionless frequency $\tilde \omega$,  the sum in (\ref{cr}) is performed with enough terms until we find the function $\tilde C(\tilde r)$ to converge within a specified numerical tolerance of 1 part in 10$^4$-10$^5$ in a given interval of positions values $\tilde r$. This means that if we keep reducing the frequency $\tilde \omega$, the correlation function does not change anymore, thus reaching the thermodynamic limit. 
 
\begin{figure}[h]
\begin{center}
\includegraphics[width=8.5cm, height=5.7cm]{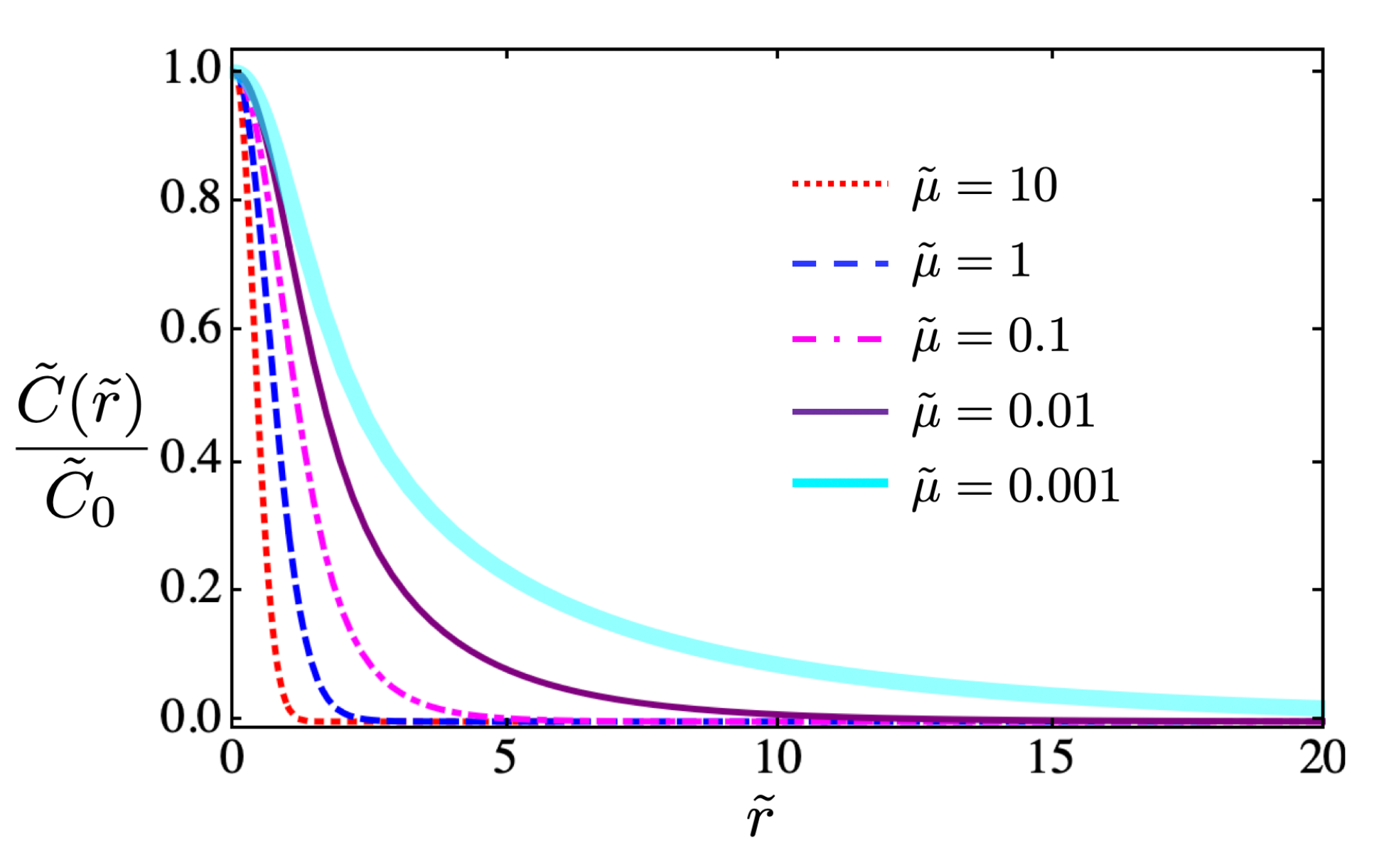}
\end{center}
\caption{Normalized density-density correlation $\tilde C(\tilde r)/\tilde C_0$, with $\tilde  C_0 = \tilde C(0)$, for several values of the chemical potential 
%
%
$\tilde \mu = 10, 1, 0.1, 0.01, 0.001$ respectively. Note the growth of the Gaussian waist as the chemical potential approaches zero.}
\label{Fig7}
\end{figure}

\begin{figure}[h]
\begin{center}
\includegraphics[width=8.5cm, height=5.7cm]{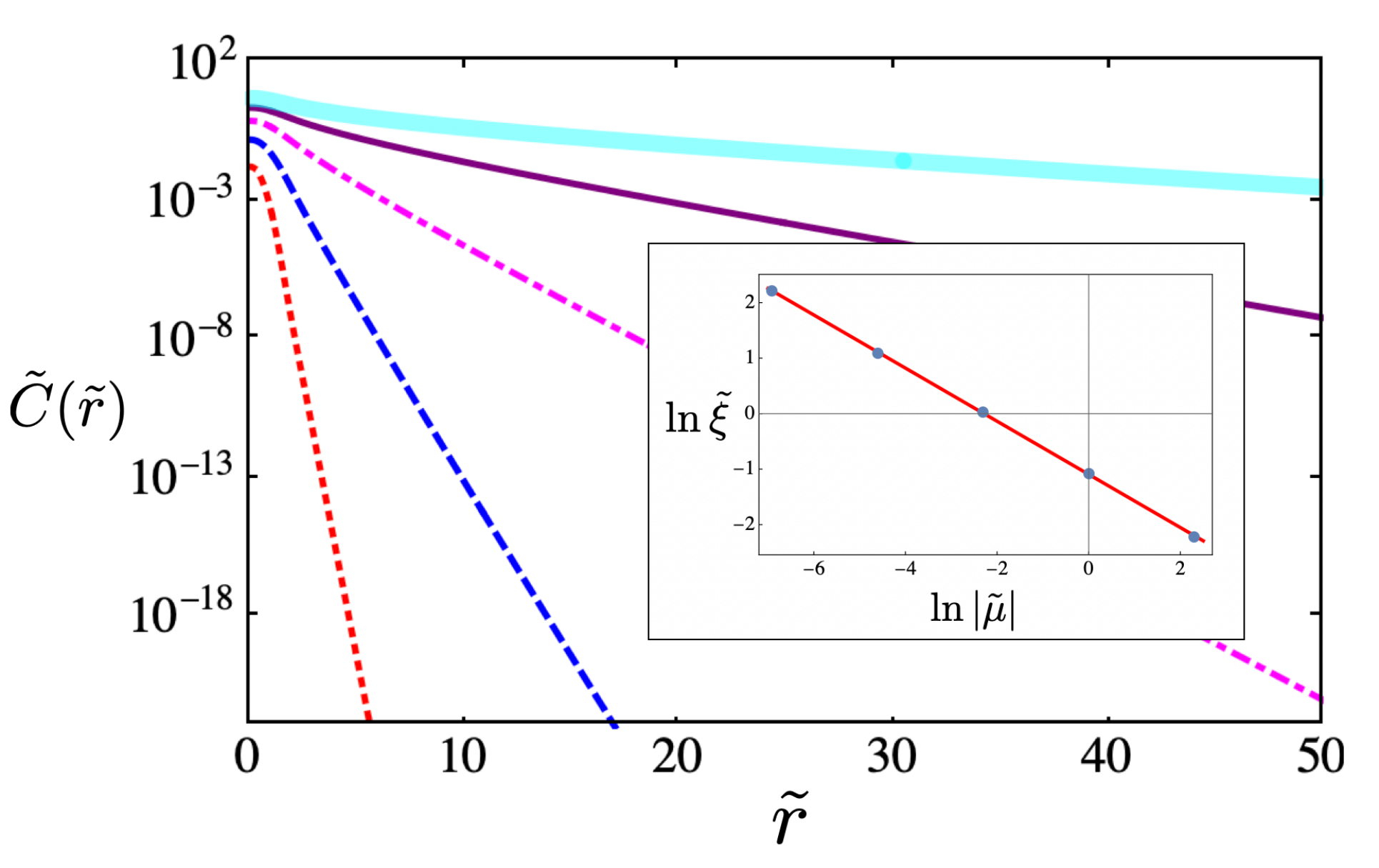}
\end{center}
\caption{Density-density correlation $\tilde C(\tilde r)$ in semi-log scale, for the same values of $\tilde \mu$ as in figure \ref{Fig7}. In the inset we show in log-log the slopes associated to the long distance behavior of $\tilde C(\tilde r)$. The red curve in the inset is the correlation length as a function of the chemical potential, the slope $\approx 0.48$ is very close to 0.5.}
\label{Fig8}
\end{figure}

The resulting correlation function $\tilde C(\tilde r)$, for five orders of magnitude  of values of the 
 dimensionless chemical potential $\tilde \mu = 10, 1, 0.1, 0.01, 0.001$, is shown in figures (\ref{Fig7}) and (\ref{Fig8}). From the former we observe that the correlation function initiates with a gaussian-like shape that becomes wider as the chemical potential becomes smaller or, equivalently, as the critical temperature is approached. However, the large distance behavior of $\tilde C(\tilde r)$ cannot be extracted from figure (\ref{Fig7}) and in figure (\ref{Fig8}) we plot the same data in a semi-log scale unveiling the  asymptotic exponential decay, that we expect to be of the form,
\begin{equation}
C(r) \sim e^{-r/\xi}
\end{equation}
with probably an accompanying algebraic decay. To each curve in figure (\ref{Fig8}) we fit a straight line in the long $\tilde r$ regime, whose slope inverses are identified as the dimensionless correlation length $\tilde \xi= \xi/(\hbar^2/mk_BT_c)^{1/2}$, and plotted in the inset against the  chemical potential $\tilde \mu$. One obtains the following numerical fit
\begin{equation}
\tilde \xi \approx 0.33\> |\tilde \mu|^{-0.48} \>.
\end{equation}
By rounding up $0.48 \approx 0.5$ and $0.33 \approx 1/2 \sqrt{2}$, and recovering the original units, one finds that such a relationship is numerically very close to
\begin{equation}
\xi = \frac{1}{2} \sqrt{\frac{\hbar^2}{2m|\mu|}} \>.\label{xi-good}
\end{equation}
First of all, we indeed find that as $\mu \to 0^{-}$, namely as BEC is approached, the correlation length diverges $\xi \to \infty$, unveiling critical density fluctuations. By appealing to the relationship (\ref{muTc}) between $T-T_c$ and $\mu$ near BEC, and using the definition of the exponent $\nu$ \cite{Stanley,Ma,Amit} $\xi \sim |T-T_c|^{-\nu_c}$, we can say that $\nu_c = 1/2$ but with logarithmic corrections.

It is of relevance to point out that the dependence of the correlation length $\xi$ on the chemical potential $\mu$ in (\ref{xi-good}) is the same as in BEC in a 3D box, where the result can be found analytically \cite{Landau,Reyes-Ayala}, however, the dependence of $\mu$ on $T-T_c$ does differ between these two cases, as detailed in the previous section. It is also of interest to recall again that experimentally, in weakly interacting gases \cite{Donner}, the critical exponent in 3D harmonic traps is $\nu \approx 0.67$ the 3D XY model prediction \cite{Campostrini,Burovski}, while theoretically, using the a Gross-Pitavskii-like theory \cite{Bezett} it has been found $\nu \approx 0.8$.

Before ending this section it is important to notice that although our analysis was restricted to the case where the  correlation function was calculated from the center of the trap towards a point separated a distance $r$, the results can be extrapolated to points located in arbitrary locations. While the correlation function will carry the presence of the harmonic confinement considering the specific spatial points where it is evaluated, one could expect the correlation length to maintain the same dependence on the chemical potential since it arises from a long distance behavior in the thermodynamic limit. 

\section{Conclusions}
\label{Conclusion}
This article presents a detailed analysis of the transition to condensation that happens in an ideal Bose gas confined by a two dimensional isotropic harmonic trap. After deriving the equation of state in terms of the proper thermodynamic variables, namely the global volume ${\cal V}= 1/\omega^2$ and its conjugate, the global pressure ${\cal P}$, we have studied the behavior of the susceptibilities to register how discontinuities and divergences signals a second order phase transition. As a key indication of such a continuous transition, the heat capacities at constant global volume $C_{{\cal V}}$ and at constant global pressure $C_{{\cal P}}$, as well as the isothermal compressibility $\kappa_T$ displayed a discontinuous and divergent behaviors as $T \to T_c^+$, respectively. By expressing the behavior near the vicinity of the critical curve $\mu=0$ and taking the leading order from the asymptotic expansions of the corresponding Bose functions, we were able to extract the critical exponents that characterize the transition, remarkably, with logarithmic corrections. These corrections have long been a topic of discussion, see Refs. \cite{Wegner,Adler,Kenna} for example. Usually those corrections arise from analysis of finite-size scaling and, in the appropriate limit, the usual scaling exponent equalities are thus modified with additional exponents associated to the logarithmic corrections \cite{Kenna}. One observation is that the logarithmic corrections in the present case are intrinsic of the system in the thermodynamic limit as we have found. And as an additional observations, we point out that the scaling hypothesis \cite{Ma,Amit} leading to the exponents equalities has been derived for homogeneous systems. This should certainly prevent us from trying to verify or find relationships among the present exponents, but clearly opens a further research topic of critical phenomena in confined inhomogeneous systems.

One of the most notable results of this investigation is the determination of the correlation length $\xi$ associated with the long-distance behavior of the density-density correlation function at the transition temperature $T \to T_c$. A meticulous numerical analysis around the critical temperature, namely as the chemical potential approaches zero $\mu \to 0^-$, allows us to discern the dependence of $\xi$ on $\mu$, which is the same as that of a homogeneous BEC in 3D. It is worth mentioning that our results foresee the dependence of $\xi$ for anisotropic ideal Bose gases, in such a case $\xi$ could include numerical factors accounting for the unequal behavior along different directions. Summarizing, we confirmed  the scale invariance of the correlation length $\xi$ as the chemical potential approaches zero $\mu \to 0^-$, and found sufficient indications of a universality class proper of the inhomogeneous harmonic confinement in 2D.

\acknowledgments{
This work was partially funded by Grant No. IN117623 from DGAPA (UNAM). MIMA acknowledges DGAPA-PAPIIT scholarship.
}

\appendix

\section{Density-density correlation function of an ideal Bose gas}
\label{AppendixB}

After substituting the creation and annihilation field operators ${\hat \Psi}^\dagger({\bar r})$ and ${\hat \Psi}({\bar r})$ in the most general expression for the density-density correlation function  $
C({\bf r} ,{\bf r'})= \langle {\hat \rho}({\bar r}) {\hat \rho}({\bar r'})\rangle - \langle{\hat \rho}({\bar r}) \rangle \langle{\hat \rho}({\bar r'})\rangle$ one obtains,
\small{
\begin{eqnarray*}
C({\bf r} ,{\bf r'}) &=& \sum_{{\bf m_1}} \sum_{{\bf m_2}} \sum_{{\bf m_3}} \sum_{{\bf m_4}} \phi_{{\bf m_1}}^*(\bf r) \phi_{{\bf m_2}}(\bf r)  \phi_{{\bf m_3}}^*(\bf r')  \phi_{{\bf m_4}}(\bf r') \nonumber \\ & \times & \left[ \langle \hat a_{\bf m_1}^\dagger \hat a_{\bf m_2}  \hat a_{\bf m_3}^\dagger \hat a_{\bf m_4} \rangle - \langle {\hat a_{\bf m_1}}^\dagger  {\hat a_{\bf m_2}} \rangle \langle {\hat a_{\bf m_3}}^\dagger  {\hat a_{\bf m_4}} \rangle \right], \nonumber
\end{eqnarray*}} 

\noindent
where $\phi_{{\bf m}}(\bf r)$ are single particle wave functions belonging to a complete basis, and the expectation value $\langle \cdot \rangle$ is taken through  the grand canonical equilibrium density matrix associated with a Bose gas having chemical potential $\mu$ and temperature $T$. To have an explicit expression for the correlation function we first consider the contribution of the expectation values with four operators  $\langle \hat a_{\bf m_1}^\dagger \hat a_{\bf m_2}  \hat a_{\bf m_3}^\dagger \hat a_{\bf m_4} \rangle$. The sums in this term can be separated as,

\small{
\begin{flalign*}
 \langle \hat a_{\bf m_1}^\dagger \hat a_{\bf m_2} & \hat a_{\bf m_3}^\dagger \hat a_{\bf m_4} \rangle  = &\\ \nonumber
&= \delta_{{\bf m_1},{\bf m_2}} \delta_{{\bf m_3},{\bf m_4}}( 1-\delta_{{\bf m_1},{\bf m_3}} )  \langle a^\dagger_{{\bf m_1}} a_{{\bf m_2}}\rangle  \langle a^\dagger_{{\bf m_3}} a_{{\bf m_4}}\rangle &\\ \nonumber
&+ \delta_{{\bf m_1},{\bf m_4}} \delta_{{\bf m_2},{\bf m_3}}( 1-\delta_{{\bf m_1},{\bf m_2}} )  \langle a^\dagger_{{\bf m_1}} a_{{\bf m_4}}\rangle \langle a_{{\bf m_2}} a^\dagger_{{\bf m_3}}\rangle &\\ \nonumber
&+ \delta_{{\bf m_1},{\bf m_2}} \delta_{{\bf m_1},{\bf m_3}} \delta_{{\bf m_1},{\bf m_4}}   \langle \hat a_{\bf m_1}^\dagger \hat a_{\bf m_2}  \hat a_{\bf m_3}^\dagger \hat a_{\bf m_4} \rangle. \nonumber
\end{flalign*}
}

\noindent
then,
\small{
\begin{flalign*}
 \sum_{{\bf m_1}} \sum_{{\bf m_2}} \sum_{{\bf m_3}} &\sum_{{\bf m_4}} \phi_{{\bf m_1}}^*({\bf r}) \phi_{{\bf m_2}}({\bf r})  \phi_{{\bf m_3}}^*({\bf r'})  \phi_{{\bf m_4}}({\bf r'}) \langle \hat a_{\bf m_1}^\dagger \hat a_{\bf m_2}  \hat a_{\bf m_3}^\dagger \hat a_{\bf m_4} \rangle = &\\ \nonumber
 & = \sum_{{\bf m_1}} \sum_{{\bf m_3}} \phi_{{\bf m_1}}^*({\bf r}) \phi_{{\bf m_1}}({\bf r})  \phi_{{\bf m_3}}^*({\bf r'})  \phi_{{\bf m_3}}({\bf r'}) \bar{n}_{{\bf m_1}} \bar{n}_{{\bf m_3}} & \\ \nonumber
 & + \sum_{{\bf m_1}} \sum_{{\bf m_2}} \phi_{{\bf m_1}}^*({\bf r}) \phi_{{\bf m_1}}({\bf r})  \phi_{{\bf m_2}}({\bf r'})  \phi_{{\bf m_2}}^*({\bf r'}) \bar{n}_{{\bf m_1}} (1+\bar{n}_{{\bf m_2}} )& \\ \nonumber
  & + \sum_{{\bf m_1}}  \phi_{{\bf m_1}}^*({\bf r}) \phi_{{\bf m_1}}({\bf r})  \phi_{{\bf m_1}}^*({\bf r'}) \phi_{{\bf m_1}}({\bf r'}) (\bar{n}_{{\bf m_1}}+2\bar{n}_{{\bf m_1}} \bar{n}_{{\bf m_1}} ), \\ \nonumber
\end{flalign*}
}

\noindent
where we have considered that $\langle {\hat a_{\bf m_j}}^\dagger  {\hat a_{\bf m_j}} \rangle = \bar{n}_{{\bf m_j}}$ and $\langle {\hat a_{\bf m_j}}^\dagger  {\hat a_{\bf m_j}}  {\hat a_{\bf m_j}}^\dagger  {\hat a_{\bf m_j}} \rangle= \bar{n}_{{\bf m_j}}+ 2 \bar{n}_{{\bf m_j}} \bar{n}_{{\bf m_j}}$. After rearranging terms and subtracting the term $\langle {\hat a_{\bf m_1}}^\dagger  {\hat a_{\bf m_2}} \rangle \langle {\hat a_{\bf m_3}}^\dagger  {\hat a_{\bf m_4}} \rangle$ one gets the final form for the correlation function $ C( {\bf r} ,{\bf r'})$,
\begin{equation*}
 C( {\bf r} ,{\bf r'})= \rho({\bf r})\delta({\bf r}-{\bf r}')+ \left| \sum_{{\bf m}} \phi_{{\bf m}}^*({\bf r})\phi_{\bf m}({\bf r}')\bar{n}_{{\bf m}} \right | ^2.
 \end{equation*}
\noindent
Since the system under study is 2D the sum over ${\bf m}$ implies a double sum, either in cartesian coordinates $(m_x,m_y)$ or in polar coordinates $(n,M)$, being $n$ a natural number and $M$ an integer.  Now, if one chooses ${\bf r}'=0$, by symmetry the correlation function can depend only on $r=|{\bf r}|$, therefore it must be true that $M=0$. Since in polar coordinates the energy levels can be written as $\epsilon_{nM}=\hbar \omega (2 n + M +1)$, this reduces to $\epsilon_{n}=\hbar \omega (2 n + 1)$, and the expression (\ref{cr}) for the correlation function is obtained.


\begin{thebibliography}{99}

\bibitem{Landau} L.D. Landau and E.M. Lifshitz, {\it Statistical Physics, Part 1} (Pergamon Press, Oxford, 1980).

\bibitem{Stanley} H. E. Stanley, {\it Introduction to Phase Transitions and Critical Phenomena} (Oxford University Press, Oxford, 1971).

\bibitem{Wilson} K. Wilson and J. Kogut, \href{https://www.sciencedirect.com/science/article/pii/0370157374900234} {Phys. Rep. {\bf 12}, 75 (1974).}

\bibitem{Ma} S. K. Ma, {\it Modern Theory of Critical Phenomena} (Benjamin, Reading, 1976).

\bibitem{Amit} D. J. Amit, {\it Field Theory, the Renormalization Group and Critical Phenomena} (Wiley, London, 1973.)

\bibitem{Fisher} M.E. Fisher, M.N. Fisher and D. Jasnow, \href{https://journals.aps.org/pra/pdf/10.1103/PhysRevA.8.1111} {Phys. Rev. A {\bf 8}, 1111 (1973).}

\bibitem{Pethick} C. J. Pethick and  H. Smith, {\it Bose–Einstein Condensation in Dilute Gases} (Cambridge University Press, Cambridge, 2008).

\bibitem{Campostrini} M. Campostrini, M. Hasenbusch, A. Pelissetto, P. Rossi E. and Vicari, \href{https://journals.aps.org/prb/abstract/10.1103/PhysRevB.63.214503} {Phys. Rev. B {\bf 63} 214503 (2001).}

\bibitem{Burovski} E. Burovski, J. Machta, N. Prokof\'ev, and B. Svistunov, \href{https://journals.aps.org/prb/abstract/10.1103/PhysRevB.74.132502} {Phys. Rev. B {\bf 74}, 132502 (2006).}

\bibitem{Lipa} J. A. Lipa, J. A. Nissen, D. A. Stricker, D. R. Swanson, and T. C. P. Chui, \href{https://journals.aps.org/prb/abstract/10.1103/PhysRevB.68.174518} {Phys. Rev. B {\bf 68}, 174518 (2003).}

\bibitem {Abrikosov} A.A. Abrikosov, L.P. Gorkov and I. E. Dzyaloshinki,  {\it Methods of Quantum Field Theory in Statistical Physics} (Dover Publications, Mineola,1978).

\bibitem{Fetter} A. L. Fetter, J. D. Walecka, {\it Quantum Theory of Many-particle systems} (McGraw-Hill International Edition, New York 1971).

\bibitem{Annett} J. F. Annet, {\it Superconductivity, Superfluids and Condensates} (Oxford Master Series in Condensed Matter Physics, Oxford University Press, Oxford 2004).

\bibitem{Reif} F. Reif, {\it Fundamentals of Statistical and Thermal Physics} (Mac Graw-Hill, New York 1965).

\bibitem{Huang} K. Huang, {\it Statistical Mechanics} Second Edition, (John Wiley and Sons, New York 1963). 

\bibitem{Pathria} R.K. Pathria, {\it Statistical Mechanics} (Ed. Butterworth-Heinemann, Oxford 2001).

\bibitem{Mermin-Wagner} N. D. Mermin and H. Wagner, \href{https://journals.aps.org/prl/abstract/10.1103/PhysRevLett.17.1133} {Phys. Rev. Lett. {\bf 17}, 1133 (1966).}

\bibitem{Hohenberg} P. C. Hohenberg, \href{https://journals.aps.org/pr/abstract/10.1103/PhysRev.158.383} {Phys. Rev. {\bf 158}, 383 (1967).}

\bibitem{Cho} Cho et al \href{https://iopscience.iop.org/article/10.1088/1367-2630/17/1/013038} {New J. Phys. {\bf 17} 013038 (2015).}

\bibitem{Bagnato} V. Bagnato and D. Kleppner \href{https://journals.aps.org/pra/abstract/10.1103/PhysRevA.44.7439} {Phys. Rev. A {\bf 44}, 7439 (1991).}

\bibitem{Mullin} W.J. Mullin \href{https://link.springer.com/article/10.1007/BF02395928} {J. Low Temp. Phys. {\bf 106}, 615 (1997).}

\bibitem{Romero-Rochin}  V. Romero-Roch\'{\i}n and V. S. Bagnato \href{https://www.scielo.br/j/bjp/a/pPxYzYD8M7bcstJSgnQFBdh/?lang=en} {Brazilian Journal of Physics, {\bf 35} 3A (2005).}

\bibitem{Brange} Brange et al. \href{https://journals.aps.org/pra/abstract/10.1103/PhysRevA.107.033324} {Phys. Rev. A {\bf 107}, 033324 (2023).}


\bibitem{Kruger} P. Kr\"uger, Z. Hadzibabic, and J. Dalibard \href{https://journals.aps.org/prl/abstract/10.1103/PhysRevLett.99.040402} {Phys. Rev. Lett. {\bf 99}, 040402 (2007).} 


\bibitem{Rajagopal} K.K. Rajagopal, P. Vignolo and M.P. Tosi \href{https://www.sciencedirect.com/science/article/pii/S0921452604008919} {Physica B: Condensed Matter {\bf 353}, 59 (2004).}

\bibitem{Tiesinga} Ch. Chin, R. Grimm, P. Julienne, and E. Tiesinga \href{https://journals.aps.org/rmp/abstract/10.1103/RevModPhys.82.1225} {Rev. Mod. Phys. {\bf 82}, 1225 (2010).}

\bibitem{Dalibard} R. Saint-Jalm, P. C. M. Castilho, \'E. Le Cerf, B. Bakkali-Hassani, J.-L. Ville, S. Nascimbene, J. Beugnon, and J. Dalibard \href{https://journals.aps.org/prx/pdf/10.1103/PhysRevX.9.021035} {Phys. Rev. X {\bf 9}, 021035 (2019).}

\bibitem{Hadzibabic} Z. Z. Hadzibabic, P. Kr\"uger, M. Cheneau, B. Battelier and J. Dalibard \href{https://www.nature.com/articles/nature04851} {Nature {\bf 441} 1118 (2006)}.

\bibitem{Fletcher} R. J. Fletcher, M. Robert-de-Saint-Vincent, J. Man, N. Navon, R. P. Smith, K. G. H. Viebahn, and Z. Hadzibabic \href{https://journals.aps.org/prl/pdf/10.1103/PhysRevLett.114.255302} {Phys. Rev. Lett. {\bf 114}, 255302 (2015).}
  
\bibitem{Sunami} S. Sunami,1, V. P. Singh, D. Garrick, A. Beregi, A. J. Barker, K. Luksch, E. Bentine, L. Mathey, and C. J. Foot \href{https://journals.aps.org/prl/pdf/10.1103/PhysRevLett.128.250402} {Phys. Rev. Lett. {\bf 128}, 250402 (2022).}

\bibitem{Singh} V. P. Singh and L. Mathey \href{https://iopscience.iop.org/article/10.1088/1367-2630/ac7d6f} {New J. Phys. {\bf 24}, 073024 (2022).} 

\bibitem{Donner} T. Donner, S. Ritter, T. Bourdel, A. \"Ottl, M. K\"ohl, and T. Esslinger \href{https://www.science.org/doi/10.1126/science.1138807} {Science {\bf 315}, 1556 (2006).} 

\bibitem{Hung} Hung et al \href{https://iopscience.iop.org/article/10.1088/1367-2630/13/7/075019}{New J. Phys. {\bf 13}, 075019 (2011).}

\bibitem{Cornell} M. H. Anderson, J.R. Ensher, M.R. Matthews, C. Wieman and E. A. Cornell, \href{https://www.science.org/doi/10.1126/science.269.5221.198} {Science {\bf 269}, 198 (1995).}

\bibitem{Ketterle} K.B.  Davis, M.O. Mewes, M.R. Andrews, N.J. van Druten, D.S. Durfee, D.M. Kurn   and W. Ketterle, \href{https://journals.aps.org/prl/abstract/10.1103/PhysRevLett.75.3969} {Phys. Rev. Lett. {\bf 75}, 3969 (1995).}

\bibitem{Dalfovo} I. Bloch, T. W. H\"ansch and T. Esslinger, \href{https://www.nature.com/articles/35003132} {Nature {\bf 403}, 166 (2000).}

\bibitem{Jin} C. A. Regal, M. Greiner, and D. S. Jin, \href{https://journals.aps.org/prl/abstract/10.1103/PhysRevLett.92.040403} {Phys. Rev. Lett. {\bf 92} 040403 (2004).}

\bibitem{Gerbier} J. Dalibard, F. Gerbier, G. Juzeliunas, and P. \"Ohberg, \href{https://journals.aps.org/rmp/abstract/10.1103/RevModPhys.83.1523} {Rev. Mod. Phys. {\bf 83} 1523 (2011).}

\bibitem{Romero-Rochin3} V. Romero-Roch\'{\i}n, \href{https://journals.aps.org/prl/pdf/10.1103/PhysRevLett.94.130601} {Phys. Rev. Lett. {\bf 94}, 130601 (2005).} 

\bibitem{Sandoval-Figueroa} N. Sandoval-Figueroa and V. Romero-Roch\'{\i}n  \href{https://journals.aps.org/pre/pdf/10.1103/PhysRevE.78.061129} {Phys. Rev. E, {\bf 78} 061129 (2008).}

\bibitem{Wegner} F. J. Wegner and E. K. Riedel, \href{https://journals.aps.org/prb/abstract/10.1103/PhysRevB.7.248}{Phys. Rev. B {\bf 7}, 248 (1973).} 

\bibitem{Adler} J. Adler and V. Privman, \href{https://iopscience.iop.org/article/10.1088/0305-4470/14/11/008} { J. Phys. A: Math. Gen. {\bf 14} L463 (1981).} 

\bibitem{Kenna} R. Kenna, \href{https://www.worldscientific.com/doi/10.1142/9789814417891_0001}{\it Order, Disorder and Criticality, Advanced Problems of Phase Transition Theory, Vol. 3, Y. Holovatch, Editor} (World Scientific, Singapore 2012), pp 1-46.

\bibitem{Vander-banda} R. R. Silva, E. A. L. Henn, K. M. F. Magalh\~aes, L. G. Marcassa, V. Romero-Rochin and V. S. Bagnato  \href{https://link.springer.com/article/10.1134/S1054660X06040244} {Las. Phys. {\bf 16}, 687 (2006).}

\bibitem{Vander-banda-2} F. J. Poveda-Cuevas, P. C. M. Castilho, E. D. Mercado-Gutierrez, A. R. Fritsch, S. R. Muniz, E. Lucioni, G. Roati, and V. S. Bagnato, \href{https://journals.aps.org/pra/abstract/10.1103/PhysRevA.92.013638} {Phys. Rev. A {\bf 92}, 013638 (2015).} 


\bibitem{Wiki} Wikipedia, Asymptotic series of the Bose-functions \href{https://en.wikipedia.org/wiki/Polylogarithm} {Wikipedia}, D. Wood  The computation of polylogarithms Technical Report 15-92*, University of Kent, Computing Laboratory, Canterbury, UK (1992).

\bibitem{Reyes-Ayala} I. Reyes-Ayala, F. J. Poveda-Cuevas and V Romero-Roch\'{\i}n \href{https://iopscience.iop.org/article/10.1088/1742-5468/ab4984/pdf} {J. Stat. Mech. 113102 (2019).} 

\bibitem{Gunton}   J. D. Gunton and M. J. Buckingham Phys. Rev. {\bf 166}, 152 (1968).

\bibitem{Bors} A. Bors \href{https://iopscience.iop.org/article/10.1088/0370-1328/87/2/301/pdf} {Proc. Phys. Soc. {\bf 87}, 343 (1966).}

\bibitem{Bezett} A. Bezzet and P.B. Blakie \href{https://journals.aps.org/pra/abstract/10.1103/PhysRevA.79.033611} {Phys. Rev. A {\bf 79}, 033611 (2009)} 

\end{thebibliography}
\end{document}